\documentclass[letterpaper,journal]{IEEEtran}
\usepackage{graphicx}
\usepackage{amsmath}
\usepackage{amssymb}
\usepackage{epsfig}
\usepackage{epstopdf}
\usepackage{amsthm}
\usepackage{afterpage}
\usepackage{amsmath,amssymb,amsfonts}
\usepackage{verbatim}
\usepackage{psfrag}
\usepackage{color}
\usepackage{cite}
\usepackage{orcidlink}
\usepackage{setspace}
\usepackage{adjustbox}
\usepackage{epsfig}
\usepackage{epstopdf}
\usepackage{footmisc}
\usepackage{fmtcount}
\usepackage{stackrel}
\usepackage{float}
\usepackage{breqn}
\usepackage{hyperref}
\usepackage{enumerate}
\usepackage{graphicx}
\usepackage{algorithm}
\usepackage{algorithmicx}
\usepackage{algpseudocode}
\usepackage{graphicx}
\usepackage[caption=false,font=footnotesize]{subfig}

\DeclareMathOperator*{\argmin}{arg\,min}
\usepackage{tabularx,booktabs}

\begin{document}

\title{Common-Transmitter Multipath-Aware TDoA Localization for Acoustic Backscatter-Enabled IoUT Networks}
\author{Ruhul Amin Khalil\orcidlink{0000-0003-4039-9901}, \IEEEmembership{Member, IEEE}
\thanks{This work was supported by the Office of the Associate Provost for Research at the United Arab Emirates University (UAEU), UAE.}
\thanks{R. A. Khalil is with the Engineering Requirement Unit (ERU), College of Engineering, United Arab Emirates University, Al-Ain 15551, UAE (e-mail: ruhulamin@uaeu.ac.ae).}
}

\maketitle

\begin{abstract}
Acoustic backscatter enables low-power Internet of Underwater Things (IoUT) nodes, but localizing passive/semi-passive tags is difficult under weak returns and surface/seabed multipath. This letter proposes a clock-free common-transmitter time-difference-of-arrival (TDoA) framework, in which one anchor interrogates the tag and synchronized receivers measure the backscattered arrivals. Receiver-side differencing cancels the unknown interrogation time, forward anchor-to-tag delay, and tag switching delay. A weighted nonlinear least-squares estimator combines signal-to-noise ratio (SNR), timing variance, and root-mean-square (RMS) delay spread. Simulations against received signal strength indicator (RSSI)-only, unweighted, SNR-weighted, and robust TDoA baselines show that at $20$ dB SNR, the root-mean-square error (RMSE) decreases from $4.71$ m to $4.13$ m, and the $5$ m success probability improves from $0.793$ to $0.835$.
\end{abstract}


\section{Introduction}
\IEEEPARstart{T}{he} Internet of Underwater Things (IoUT) connects underwater sensors, anchors, vehicles, and surface gateways for environmental monitoring, asset inspection, disaster warning, marine data collection, and security applications~\cite{jahanbakht2021iout}. In these applications, localization is essential for geo-referencing sensed data, guiding mobile nodes, and supporting routing or scheduling decisions~\cite{khalil2021bayesian}. However, underwater localization is difficult because radio-frequency and satellite signals are not directly available below the water surface.

Acoustic waves are the main carrier for underwater communication and positioning, but they suffer from low propagation speed, limited bandwidth, long latency, Doppler sensitivity, frequency-dependent attenuation, and surface/seabed multipath~\cite{watson2020uuv,liu2024beacon,khalil2020toward}. Existing solutions include long-/short-baseline systems, beacon-assisted navigation, inertial/dead-reckoning fusion, and autonomous underwater vehicle (AUV)-assisted localization~\cite{alexandris2024positioning}. Time-based and anchor-assisted schemes remain attractive, but their accuracy is affected by synchronization, geometry, multipath, and energy consumption.

Acoustic backscatter communication is promising for low-power IoUT localization because a passive/semi-passive node modulates and reflects an incident acoustic waveform rather than actively transmitting a high-power packet~\cite{jang2019underwater}. Piezo-acoustic backscatter, piezoelectric metamaterial tags, Van Atta acoustic backscatter, and acoustic identification tags have improved bandwidth, directivity, link range, and identifiability~\cite{eid2023vab}. Still, localization is challenging because the returned signal is weak, multipath-contaminated, and the tag usually cannot maintain a precise synchronized clock~\cite{khalil2026semantic}.

Recent underwater backscatter localization has mainly considered vehicle-centered operation. For instance,~\cite{zhou2025precise} used seabed acoustic identification tags and an acoustic-intensity-vector array for AUV localization. This is useful for local AUV guidance, but the reference is tied to a moving platform and array-specific hardware. In contrast, fixed IoUT anchors provide stable infrastructure and allow the tag to remain simple, low-power, and clock-free.

Unlike conventional multistatic TDoA and passive acoustic source localization, the unknown IoUT node in this work is not an active synchronized emitter; it only reflects and modulates the waveform transmitted by a common interrogating anchor. Hence, the raw backscatter arrival contains the unknown interrogation time, common forward anchor-to-tag delay, and tag switching delay. Receiver-side differencing cancels these backscatter-specific timing terms, enabling clock-free localization of passive/semi-passive acoustic backscatter IoUT nodes.

The main contributions are as follows:
\begin{itemize}
    \setlength{\itemsep}{0pt}
    \setlength{\topsep}{0pt}
    \setlength{\parsep}{0pt}
    \item We propose a common-transmitter acoustic-backscatter TDoA model that converts the cascaded interrogation--reflection timing relation into a receiver-side range-difference problem.
    \item We develop a multipath-aware weighted nonlinear least-squares estimator using SNR, timing-noise variance, and RMS delay spread to suppress unreliable receiver pairs.
    \item We compare the proposed method with RSSI-only localization \cite{li2026rss}, unweighted TDoA \cite{mandic2024tdoa}, SNR-weighted TDoA \cite{zhu2021nonstationary}, robust TDoA \cite{abrar2025robust}, and a noise-only Fisher-information CRLB benchmark.
\end{itemize}

\section{System Model}\label{systemmodel}
\begin{figure}[!t]
    \centering
    \includegraphics[width=\columnwidth]{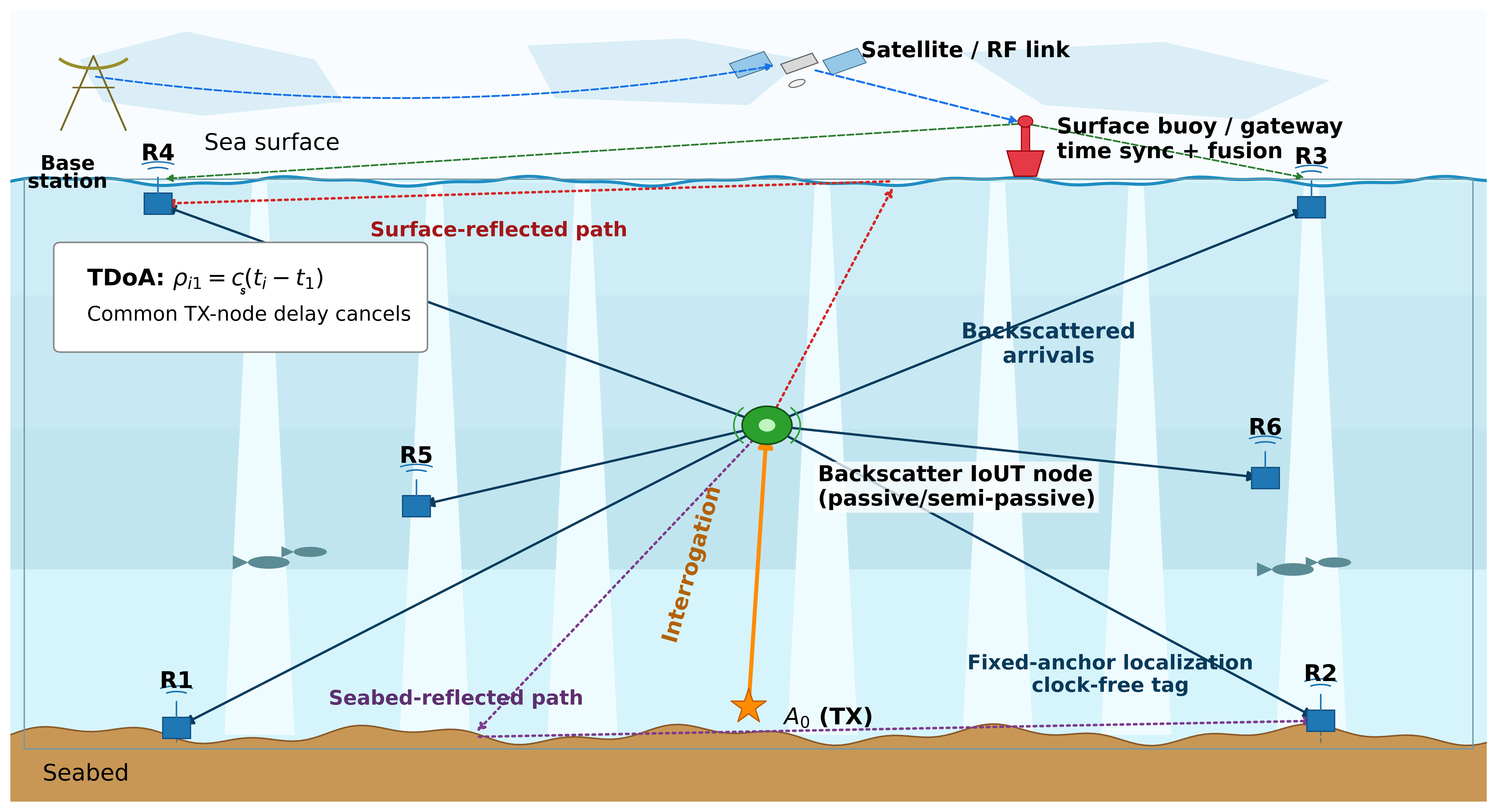}
    \caption{Proposed acoustic backscatter-enabled IoUT localization network.}
    \label{fig:system_model}
\end{figure}

The proposed model considers fixed infrastructure anchors and a passive/semi-passive backscatter IoUT node as the unknown target. Fig.~\ref{fig:system_model} illustrates the fixed-anchor acoustic backscatter localization network.

\subsection{Network Architecture}
Consider one fixed interrogating anchor $A_0$ and $M$ synchronized receiving anchors $R_i$, $i=1,\ldots,M$, synchronized through a surface buoy, cabled reference, high-stability clocks, or periodic acoustic synchronization \cite{mandic2024tdoa}. Their known positions are
\begin{equation}
    \mathbf{a}_i=[x_i,y_i,z_i]^T,\quad i=0,1,\ldots,M,
\end{equation}
where $\mathbf{a}_0$ is the position of $A_0$, and $\mathbf{a}_i$, $i=1,\ldots,M$, is the position of receiver $R_i$. The unknown node position is
\begin{equation}
    \mathbf{p}=[x,y,z]^T .
\end{equation}
Known-depth localization requires at least three receiving anchors, while full 3D localization requires at least four because $M-1$ independent TDoA observations are available. These are only identifiability minima; redundant receivers improve FIM conditioning, reduce geometric dilution, and support multipath-aware down-weighting.

Let
\begin{equation}
    d_0(\mathbf{p})=\|\mathbf{p}-\mathbf{a}_0\|,\quad
    d_i(\mathbf{p})=\|\mathbf{p}-\mathbf{a}_i\|,
\end{equation}
where $d_0$ is the forward interrogation distance from $A_0$ to the tag, and $d_i$, $i=1,\ldots,M$, is the backscatter return distance from the tag to $R_i$. The cascaded interrogation--backscatter propagation distance through $R_i$ is
\begin{equation}
    D_i(\mathbf{p})=d_0(\mathbf{p})+d_i(\mathbf{p}).
\end{equation}
The anchor transmits a known waveform, the tag modulates its reflection, and the receivers forward timing and channel-quality estimates to the gateway.

\subsection{Underwater Backscatter Channel}
The received signal at $R_i$ experiences cascaded acoustic spreading, absorption, and multipath over the interrogation and backscatter paths \cite{ghaffarivardavagh2020uwb}. A compact baseband model is
\begin{equation}
\small
    y_i(t)=\sum_{\ell=0}^{L_i-1}\alpha_{i,\ell}q(t-\tau_{i,\ell})+v_i(t),\quad q(t)=b(t)s(t),
    \label{eq:channel}
\end{equation}
where $s(t)$ is the interrogation waveform, $b(t)$ is the tag response, $\ell=0,\ldots,L_i-1$ indexes the resolvable components at $R_i$, $L_i$ is their total number, $\alpha_{i,\ell}$ and $\tau_{i,\ell}$ are the gain and delay of the $\ell$-th component, and $v_i(t)$ is ambient noise. The earliest significant matched-filter peak above the threshold is treated as the direct component when resolvable; otherwise, unresolved multipath increases the PDP spread and reduces the corresponding reliability weight.

For carrier frequency $f$, the approximate cascaded acoustic path loss in dB is
\begin{equation}
    \mathcal{L}_i(f)=10\kappa\log_{10}\!\left(d_0d_i\right)+\alpha(f)(d_0+d_i),
    \label{eq:pathloss}
\end{equation}
where $\kappa$ is the spreading factor and $\alpha(f)$ is the absorption coefficient in dB/m. The RMS delay spread from the power-delay profile is
\begin{equation}
\small
\tau_{\mathrm{rms},i}=\sqrt{\frac{\sum_{\ell}|\alpha_{i,\ell}|^2(\tau_{i,\ell}-\bar{\tau}_i)^2}{\sum_{\ell}|\alpha_{i,\ell}|^2}},\quad
    \bar{\tau}_i=\frac{\sum_{\ell}|\alpha_{i,\ell}|^2\tau_{i,\ell}}{\sum_{\ell}|\alpha_{i,\ell}|^2}.
    \label{eq:rmsdelay}
\end{equation}
The terms in \eqref{eq:pathloss} model distance- and frequency-dependent acoustic loss, while \eqref{eq:rmsdelay} provides the multipath-dispersion reliability metric used in the proposed TDoA weighting \cite{zhu2021nonstationary,graupe2022framework}.

\section{Common-Transmitter Backscatter TDoA Model}\label{backscattermodel}
\subsection{Arrival-Time Model}
Let $t_0$ be the unknown interrogation time, $c_s$ be the underwater sound speed, and $\delta_b$ be the unknown backscatter switching delay. The arrival time estimated at receiver $R_i$ is
\begin{equation}
\small
\label{eq:arrival}
t_i=t_0+\frac{d_0(\mathbf{p})}{c_s}+\delta_b+
\frac{d_i(\mathbf{p})}{c_s}+n_i+e_i^{\mathrm{mp}} .
\end{equation}
where $n_i$ is timing-estimation noise and $e_i^{\mathrm{mp}}$ is multipath-induced timing bias. Since the same transmitter $A_0$ interrogates the tag, $t_0$, $d_0(\mathbf{p})/c_s$, and $\delta_b$ are common to all receivers.

Choosing $R_1$ as the reference receiver, the TDoA measurement is
\begin{equation}
\label{eq:tdoa}
    \Delta t_{i1}=t_i-t_1,\quad i=2,\ldots,M.
\end{equation}
Substituting \eqref{eq:arrival} into \eqref{eq:tdoa} gives
\begin{equation}
\label{eq:tdoa_expand}
    \Delta t_{i1}=\frac{d_i(\mathbf{p})-d_1(\mathbf{p})}{c_s}
+\tilde{n}_{i1}+\tilde{e}_{i1}^{\mathrm{mp}},
\end{equation}
where $\tilde{n}_{i1}=n_i-n_1$ and $\tilde{e}_{i1}^{\mathrm{mp}}=e_i^{\mathrm{mp}}-e_1^{\mathrm{mp}}$. Thus, the tag does not need an absolute clock, i.e., it does not need to know the interrogation time or maintain synchronization with the anchor network.

\subsection{Range-Difference Observation}
For range-difference, multiplying \eqref{eq:tdoa_expand} by $c_s$ yields
\begin{equation}
\label{eq:range_difference}
    \rho_{i1}=c_s\Delta t_{i1}=d_i(\mathbf{p})-d_1(\mathbf{p})+\xi_{i1},
\end{equation}
where $\xi_{i1}=c_s(\tilde{n}_{i1}+\tilde{e}_{i1}^{\mathrm{mp}})$ is the range-domain error, i.e., the TDoA timing error converted into distance units.
\begin{equation}
\label{eq:sigma_rho}
    \sigma_{\rho,i1}^{2}=c_s^2\left(\sigma_{t,i}^{2}+\sigma_{t,1}^{2}\right),\quad
    \sigma_{t,i}^{2}=\frac{\eta_t^2}{B^2\gamma_i},
\end{equation}
where $B$ is the effective bandwidth, $\gamma_i$ is the received SNR, and $\eta_t$ is a timing-estimation constant.

\section{Multipath-Aware Weighted TDoA Localization}\label{weightedTDoA}

\subsection{Measurement Reliability Weight}
Conventional TDoA either weights all receiver pairs equally or relies only on SNR \cite{mandic2024tdoa}. In underwater backscatter links, reliability also depends on the temporal spread of the power-delay profile. Thus, the proposed weight follows an approximate inverse-uncertainty principle: $\sigma_{\rho,i1}^{2}$ captures the SNR/bandwidth-dependent timing-noise variance, while $c_s^2(\hat{\tau}_{\mathrm{rms},i}^{2}+\hat{\tau}_{\mathrm{rms},1}^{2})$ penalizes receiver pairs with large multipath delay spread. We define
\begin{equation}
\label{eq:weight}
    w_{i1}=\frac{\bar{\gamma}_{i1}}{\sigma_{\rho,i1}^{2}+\beta c_s^2\left(\hat{\tau}_{\mathrm{rms},i}^{2}+\hat{\tau}_{\mathrm{rms},1}^{2}\right)+\epsilon},
\end{equation}
where $\beta$ controls the multipath penalty and $\epsilon>0$ avoids division by zero. Here, $w_{i1}$ is the reliability weight of the TDoA pair $(R_i,R_1)$, and $\hat{\tau}_{\mathrm{rms},i}$ is the RMS delay spread estimated from the thresholded PDP at receiver $R_i$. The normalized pairwise SNR reliability is
\begin{equation}
\label{eq:snrweight}
    \bar{\gamma}_{i1}=\frac{\min(\gamma_i,\gamma_1)}{\max_m \gamma_m+\epsilon}.
\end{equation}
The coefficient $\beta$ scales the PDP-based multipath penalty: $\beta=0$ reduces the method to SNR/timing-variance weighting, while larger $\beta$ increasingly down-weights receiver pairs with larger RMS delay spread. Hence, a pair receives high confidence only when both receivers observe strong and temporally compact backscatter responses.

\begin{algorithm}[t!]
\caption{Multipath-Aware Weighted TDoA Localization.}
\label{alg:maw_tdoa}
\scriptsize
\begin{algorithmic}[1]
\Require $\mathbf{a}_0$, $\{\mathbf{a}_i\}_{i=1}^{M}$, $c_s$, reference $R_1$, waveform $s(t)$
\Ensure Estimated node position $\hat{\mathbf{p}}$
\State $A_0$ transmits the interrogation waveform $s(t)$.
\State The IoUT node backscatters and modulates $s(t)$ with its identifier.
\For{$i=1$ to $M$}
    \State Receiver $R_i$ estimates $\hat{t}_i$, $\gamma_i$, and $\hat{\tau}_{\mathrm{rms},i}$.
\EndFor
\For{$i=2$ to $M$}
    \State Form $\Delta\hat{t}_{i1}=\hat{t}_i-\hat{t}_1$ and $\rho_{i1}=c_s\Delta\hat{t}_{i1}$.
    \State Compute $w_{i1}$ using \eqref{eq:weight}.
\EndFor
\State Solve \eqref{eq:wnls} using \eqref{eq:gn}.
\State \Return $\hat{\mathbf{p}}$
\end{algorithmic}
\end{algorithm}

\subsection{Weighted Least-Squares Estimator and CRLB}
The range-difference residual is
\begin{equation}
\label{eq:residual}
    r_{i1}(\mathbf{p})=\rho_{i1}-\left[d_i(\mathbf{p})-d_1(\mathbf{p})\right].
\end{equation}
The proposed weighted nonlinear least-squares (WNLS) estimate is obtained by solving
\begin{equation}
\label{eq:wnls}
\hat{\mathbf{p}}=\argmin_{\mathbf{p}\in\Omega}\sum_{i=2}^{M}w_{i1}r_{i1}^{2}(\mathbf{p}),
\end{equation}
where $\Omega$ is the feasible underwater region. The Jacobian of $h_{i1}(\mathbf{p})=d_i(\mathbf{p})-d_1(\mathbf{p})$ is
\begin{equation}
\label{eq:jacobian}
\small
    \mathbf{J}_{i1}=\frac{(\mathbf{p}-\mathbf{a}_i)^T}{\|\mathbf{p}-\mathbf{a}_i\|}-\frac{(\mathbf{p}-\mathbf{a}_1)^T}{\|\mathbf{p}-\mathbf{a}_1\|}.
\end{equation}
Since \eqref{eq:wnls} is nonlinear in $\mathbf{p}$, it is solved iteratively using a damped Gauss--Newton method. At iteration $k$, the residuals and Jacobian are evaluated at the current estimate, and the position increment is
\begin{equation}
\label{eq:gn}
    \small\Delta\mathbf{p}^{(k)}=\left(\mathbf{J}^{T}\mathbf{W}\mathbf{J}+\lambda\mathbf{I}\right)^{-1}\mathbf{J}^{T}\mathbf{W}\mathbf{r},
\end{equation}
where $\mathbf{W}=\mathrm{diag}(w_{21},\ldots,w_{M1})$, $\lambda\geq0$, and $\mathbf{r}$ stacks the residuals.

For benchmarking, let $\boldsymbol{\Sigma}_{\rho}$ be the noise-only covariance of the range-difference vector and $\mathbf{H}$ be the Jacobian in \eqref{eq:jacobian}. The Fisher information matrix (FIM) and RMSE lower bound are
\begin{equation}
\small
\label{eq:fim}
    \mathbf{F}(\mathbf{p})=\mathbf{H}^{T}\boldsymbol{\Sigma}_{\rho}^{-1}\mathbf{H},\quad \mathrm{CRLB}_{\mathrm{rmse}}=\sqrt{\mathrm{tr}\!\left(\mathbf{F}^{-1}(\mathbf{p})\right)}.
\end{equation}
The CRLB in \eqref{eq:fim} is a noise-only lower bound on unbiased TDoA errors for reference. Under multipath, delayed surface/seabed components introduce bias, so the practical error includes both variance and bias terms; the proposed weighting aims to reduce this additional bias by down-weighting high-delay-spread receiver pairs.

In practice, $\hat{t}_i$ is extracted from the matched-filter/correlation peak of the known interrogation waveform and tag identifier, $\gamma_i$ from the peak-to-noise power ratio, and $\hat{\tau}_{\mathrm{rms},i}$ from the thresholded power-delay profile. Residual clock drift, sound-speed mismatch, and tag-response variation appear as timing, SNR, or delay-spread uncertainty and are therefore reflected in the proposed reliability weight. Algorithm~\ref{alg:maw_tdoa} summarizes the procedure.

\begin{figure*}[!t]
    \centering

    \subfloat[3D TDoA hyperbolic constraints.]{
        \includegraphics[width=0.35\textwidth]{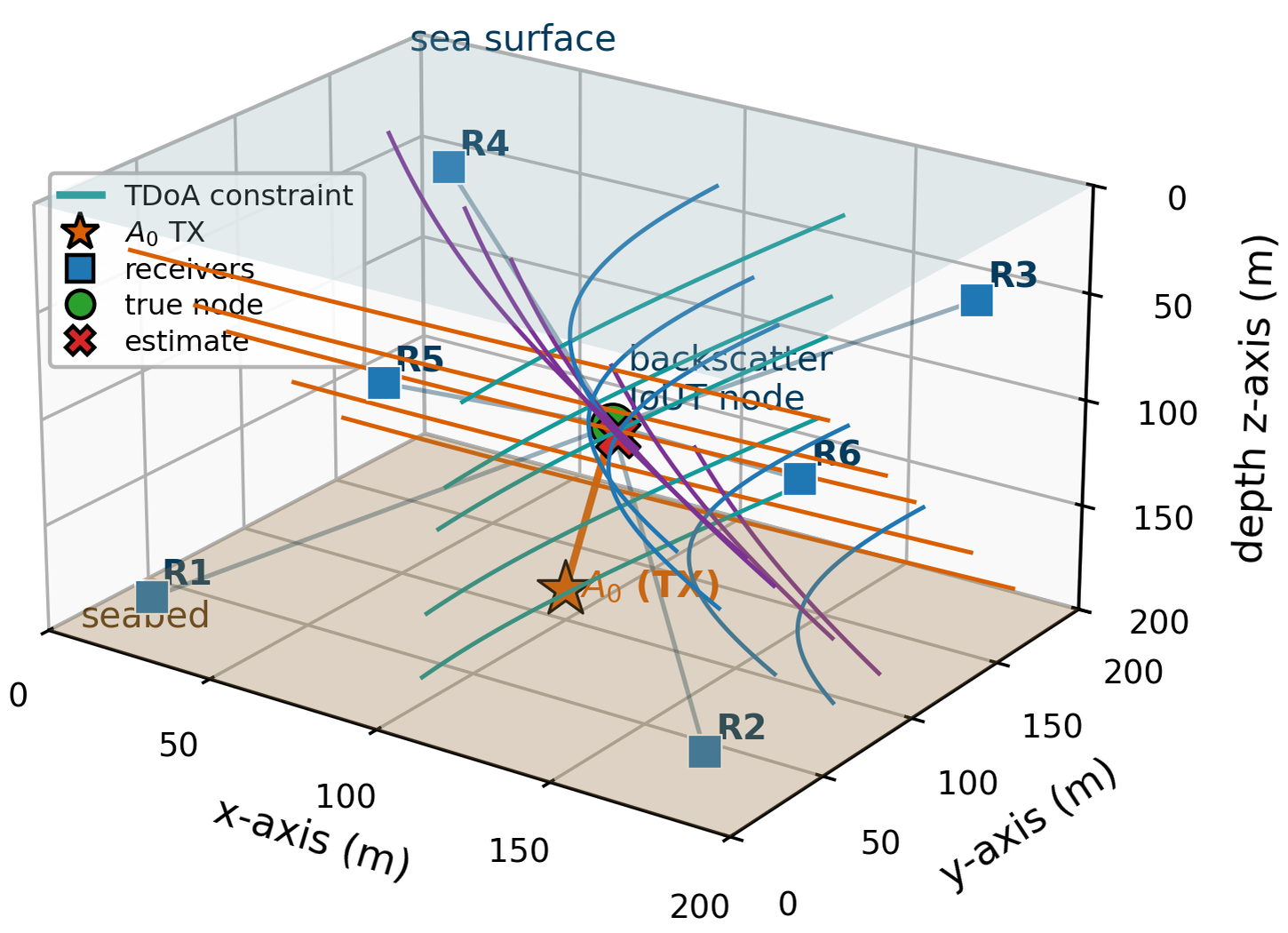}
        \label{fig:3d_tdoa_constraints}
    }
    \hspace{1.5cm}
    \subfloat[Weighted TDoA objective surface.]{
        \includegraphics[width=0.35\textwidth]{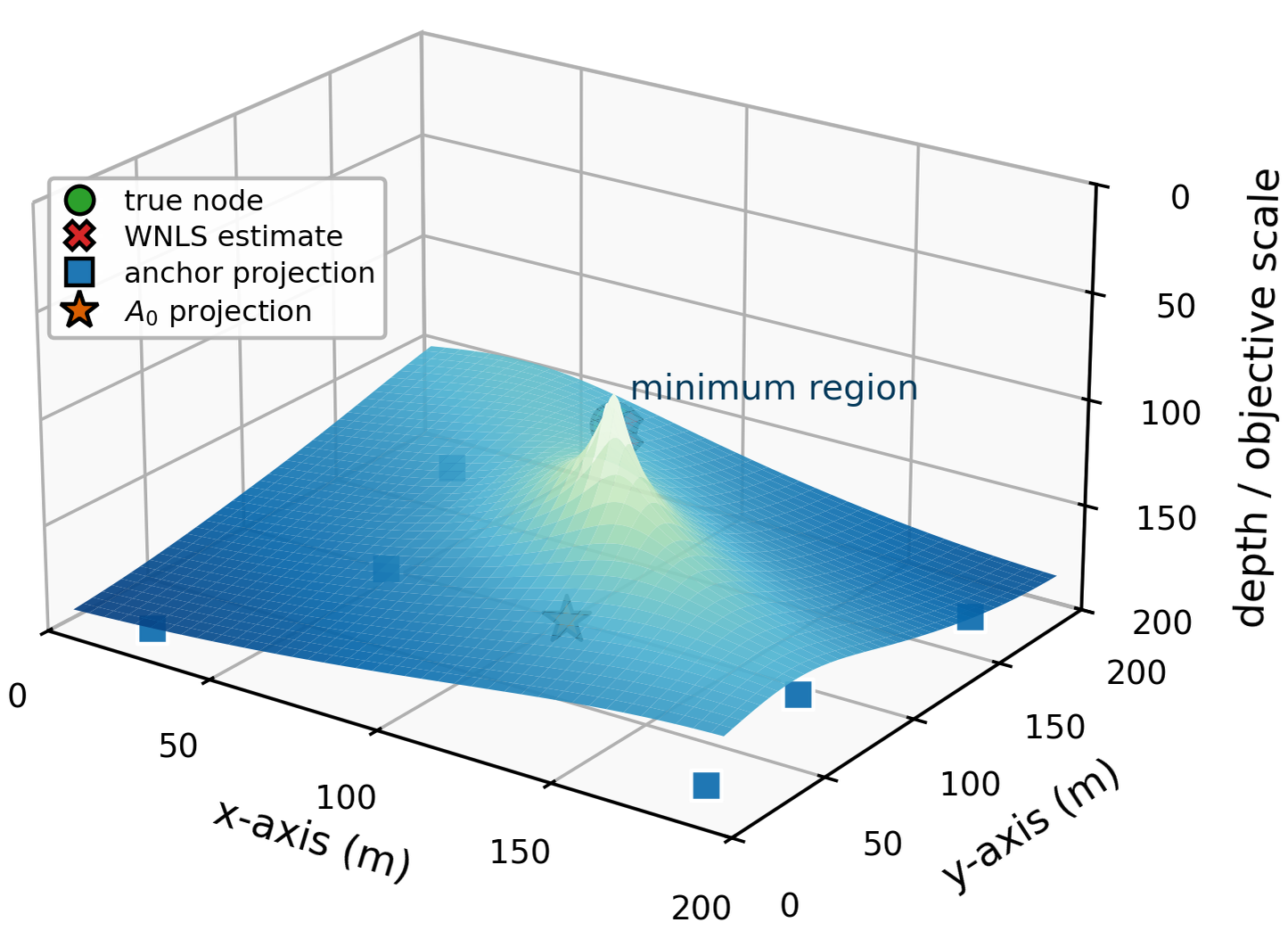}
        \label{fig:3d_cost_surface}
    }

    \vspace{0.3em}

    \subfloat[Multipath-aware receiver reliability with PDP inset.]{
        \includegraphics[width=0.35\textwidth]{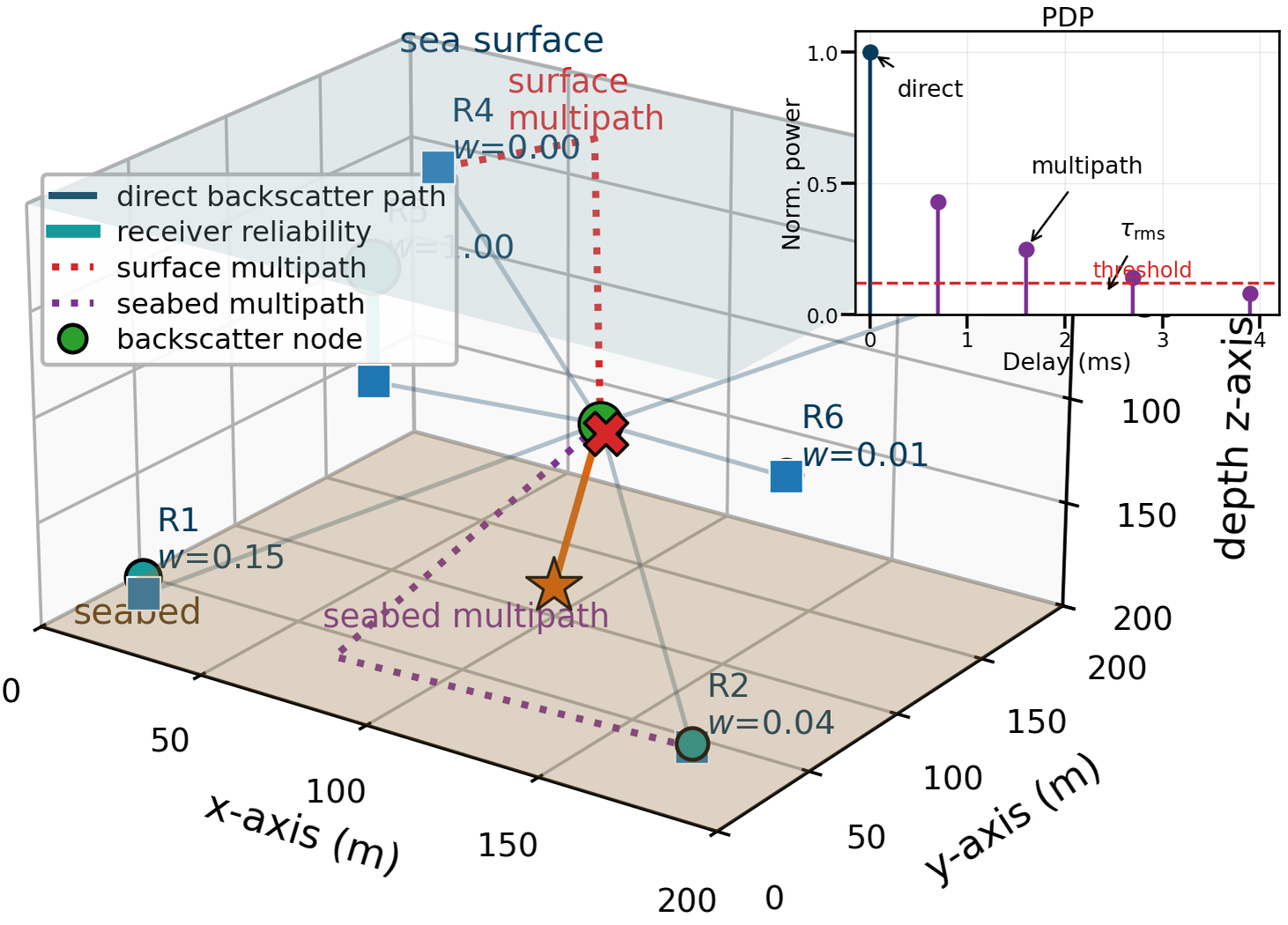}
        \label{fig:3d_reliability}
    }
    \hspace{1.5cm}
    \subfloat[Monte Carlo localization estimates.]{
        \includegraphics[width=0.35\textwidth]{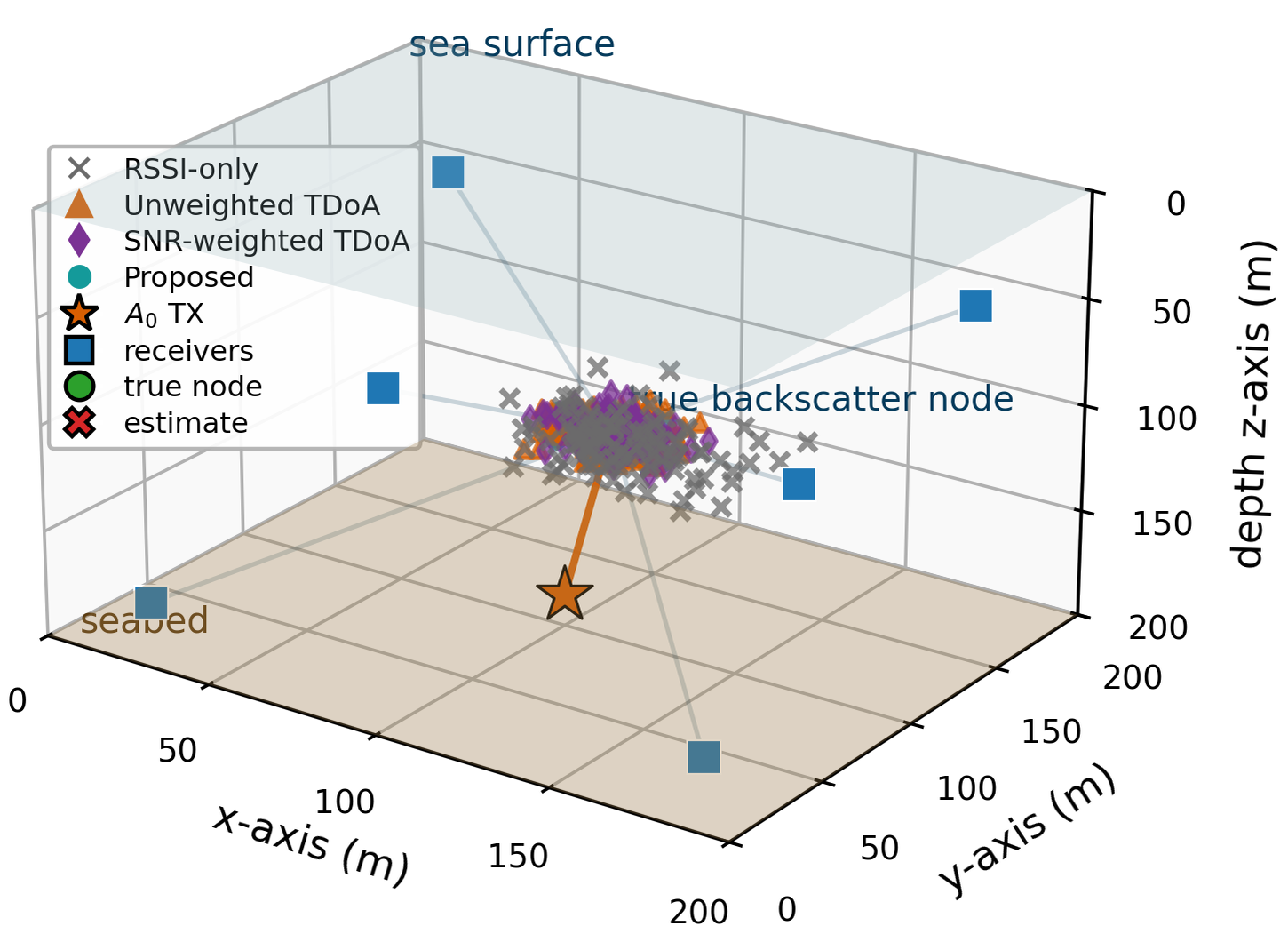}
        \label{fig:3d_mc_clouds}
    }

    \caption{3D visualization of the proposed common-transmitter acoustic-backscatter TDoA localization method in a $200\,\mathrm{m}\times200\,\mathrm{m}\times200\,\mathrm{m}$ underwater region.}
    \label{fig:3d_results}
\end{figure*}

\begin{figure*}[!t]
    \centering

    \subfloat[RMSE versus SNR.]{
        \includegraphics[width=0.315\textwidth]{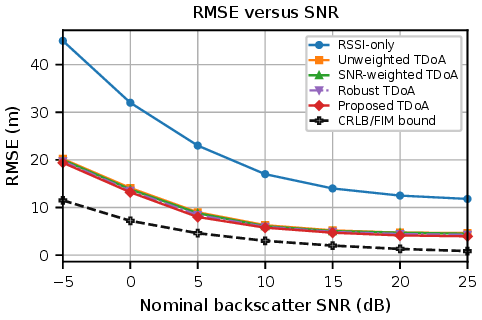}
        \label{fig:rmse_snr}
    }
    \hfill
    \subfloat[RMSE versus multipath delay spread.]{
        \includegraphics[width=0.315\textwidth]{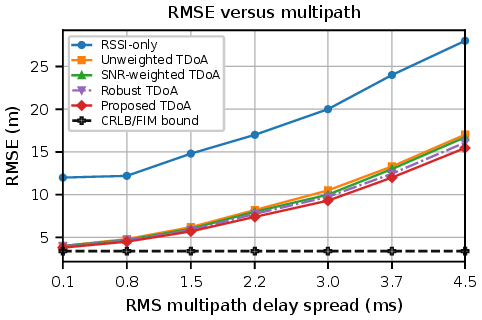}
        \label{fig:rmse_multipath}
    }
    \hfill
    \subfloat[RMSE versus number of receiving anchors.]{
        \includegraphics[width=0.315\textwidth]{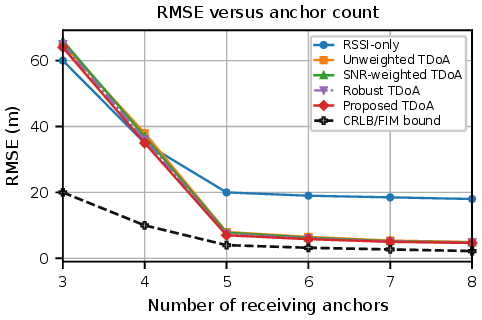}
        \label{fig:rmse_anchors}
    }

    \vspace{0.5em}

    \subfloat[Localization-error CDF.]{
        \includegraphics[width=0.315\textwidth]{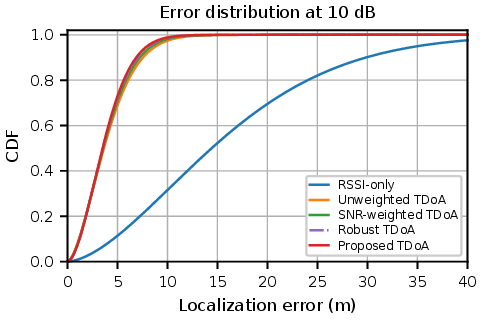}
        \label{fig:error_cdf}
    }
    \hfill
    \subfloat[RMSE versus average node--anchor distance.]{
        \includegraphics[width=0.315\textwidth]{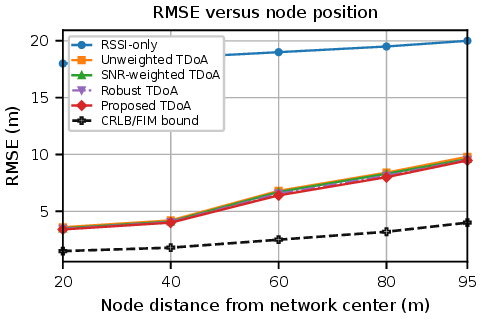}
        \label{fig:rmse_distance}
    }
    \hfill
    \subfloat[Localization success probability versus SNR.]{
        \includegraphics[width=0.315\textwidth]{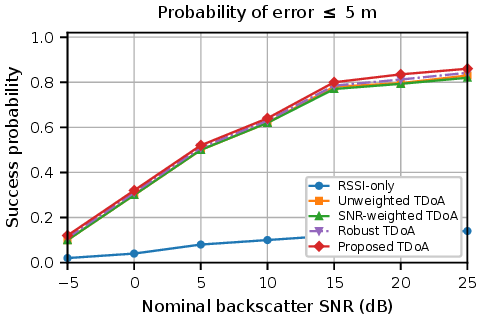}
        \label{fig:success_snr}
    }

    \caption{Performance comparison of the proposed multipath-aware weighted TDoA localization method against RSSI-only localization, unweighted TDoA, SNR-weighted TDoA, robust TDoA, and the noise-only CRLB/FIM benchmark.}
    \label{fig:simulation_results_all}
\end{figure*}

\section{Simulation Results and Discussion}\label{results}
\subsection{Simulation Setup}
A shallow-water IoUT region of $200\,\mathrm{m}\times200\,\mathrm{m}\times200\,\mathrm{m}$ is considered. Unless otherwise stated, six receiving anchors are used as a redundant default deployment, $c_s=1500\,\mathrm{m/s}$, and the node depth is assumed known for horizontal localization. The default anchor coordinates in meters are $A_0=(100,100,185)$, $R_1=(20,20,185)$, $R_2=(180,25,185)$, $R_3=(180,175,50)$, $R_4=(20,180,50)$, $R_5=(45,100,110)$, and $R_6=(170,95,100)$; $R_7=(100,45,160)$ and $R_8=(100,170,60)$ are added in anchor-count tests. Although three receivers are sufficient for known-depth localization and four for full 3D identifiability, the six-receiver setup provides additional TDoA constraints, improving geometry and multipath robustness.

\begin{table}[H]
\centering
\caption{Simulation Parameters.}
\label{tab:simparams}
\scriptsize
\begin{tabular}{@{}ll@{}}
\toprule
Parameter & Value/Model \\
\midrule
Region, trials & $200\times200\times200\,\mathrm{m}^3$, $N_{\mathrm{mc}}=1000$ \\
Carrier/bandwidth & $f_c=24$ kHz, $B=2$ kHz \\
Path loss & spreading $\kappa=1.5$, absorption $\alpha(f_c)=0.005$ dB/m \\
SNR range & received backscatter SNR, $-5$ to $25$ dB \\
Timing noise & $\sigma_{t,i}=\eta_t/(B\sqrt{\gamma_i})$, $\eta_t=10$ \\
Multipath & $\tau_{\mathrm{rms}}=0.1$--$4.5$ ms; bias variance $\propto c_s^2\tau_{\mathrm{rms}}^2$ \\
Estimator & centroid initialization, $30$ iterations, $\|\Delta\mathbf{p}\|<10^{-3}$ m \\
Weights & $\beta=0.7$, $\epsilon=10^{-9}$ \\
\bottomrule
\end{tabular}
\end{table}

Table~\ref{tab:simparams} summarizes the Monte Carlo setup. The value $\beta=0.7$ was chosen as a moderate multipath penalty based on sensitivity tests: small $\beta$ approaches SNR-weighted TDoA, while large $\beta$ may reject useful, mildly dispersed links.

The stochastic channel model enables reproducible estimator-level validation under controlled conditions of SNR, timing noise, and RMS delay spread. More realistic sound-speed variation, surface motion, seabed roughness, Doppler, refraction, tag-response variability, and non-Gaussian timing outliers can be incorporated through Bellhop/ray-tracing outputs or measured power-delay profiles.

The backscatter SNR and RMS delay spread are varied in the tests. Timing errors are SNR-dependent Gaussian, while multipath bias variance increases with delay spread. For $N_{\mathrm{mc}}$ trials, $\mathrm{RMSE}=\sqrt{N_{\mathrm{mc}}^{-1}\sum_n\|\hat{\mathbf{p}}_n-\mathbf{p}_n\|^2}$ and $P_s(e_{\mathrm{th}})=N_{\mathrm{mc}}^{-1}\sum_n\mathbb{I}(\|\hat{\mathbf{p}}_n-\mathbf{p}_n\|\leq e_{\mathrm{th}})$, with $e_{\mathrm{th}}=5$ m.

To provide a broader benchmark evaluation, five practical algorithms are considered: RSSI-only \cite{li2026rss}, unweighted TDoA \cite{mandic2024tdoa}, SNR-weighted TDoA \cite{zhu2021nonstationary}, robust TDoA \cite{abrar2025robust}, and the proposed method. SNR-weighted TDoA is obtained by setting $\beta=0$ in \eqref{eq:weight}, whereas the proposed method uses $\beta>0$ for delay-spread-aware weighting. Robust TDoA suppresses large residuals but does not exploit SNR/RMS-delay-spread information before optimization. The noise-only CRLB/FIM bound in \eqref{eq:fim} is included as a theoretical reference \cite{khalil2021bayesian}.

\subsection{Performance Comparisons}
Fig.~\ref{fig:3d_results} illustrates the common-transmitter backscatter TDoA process in the $200\,\mathrm{m}$ cubic region. Fig.~\ref{fig:3d_reliability} includes a PDP inset showing the direct peak, delayed multipath, detection threshold, and RMS delay spread used in \eqref{eq:rmsdelay}.The TDoA constraints follow underwater acoustic TDoA geometry \cite{mandic2024tdoa}, while the reliability panel reflects the SNR/RMS-delay-spread weighting motivated by underwater multipath dispersion \cite{zhu2021nonstationary}. The Monte Carlo cloud shows a tighter estimate cluster for the proposed method than RSSI-only localization \cite{li2026rss}, unweighted TDoA \cite{mandic2024tdoa}, and SNR-weighted TDoA \cite{zhu2021nonstationary}.

Fig.~\ref{fig:simulation_results_all} compares RSSI-only localization \cite{li2026rss}, unweighted TDoA \cite{mandic2024tdoa}, SNR-weighted TDoA \cite{zhu2021nonstationary}, robust TDoA \cite{abrar2025robust}, the proposed method, and the CRLB/FIM benchmark \cite{khalil2021bayesian}. The newly added robust TDoA baseline suppresses large multipath-induced residuals, but unlike the proposed method, it does not exploit SNR/RMS-delay-spread information prior to localization. At $10$ dB, the proposed method achieves about $5.76$ m RMSE, compared with $6.29$ m, $6.08$ m, and $5.93$ m for unweighted~\cite{mandic2024tdoa}, SNR-weighted~\cite{zhu2021nonstationary}, and robust TDoA~\cite{abrar2025robust}, respectively. At $20$ dB, it reaches about $4.13$ m, while the three TDoA baselines remain near $4.72$ m, $4.71$ m, and $4.43$ m. Thus, relative to SNR-weighted TDoA, the proposed method provides about $12.3\%$ RMSE reduction at $20$ dB.

Under severe multipath, the benefit of delay-spread-aware weighting becomes clearer. At $\tau_{\mathrm{rms}}=4.5$ ms, the proposed method gives about $15.46$ m RMSE, while unweighted TDoA \cite{mandic2024tdoa}, SNR-weighted TDoA \cite{zhu2021nonstationary}, and robust TDoA \cite{abrar2025robust} give about $17.02$ m, $16.72$ m, and $16.10$ m, respectively. This shows that robust residual suppression helps under multipath, but the proposed method further improves accuracy by down-weighting unreliable receiver pairs before WNLS fitting. With eight receivers, the proposed method achieves about $4.65$ m RMSE, compared with $4.95$ m, $4.85$ m, and $4.74$ m for the three TDoA baselines. This supports using six default receivers, since redundant anchors improve conditioning and provide additional measurements to down-weight unreliable links.

The CDF, distance, and success-probability results further support the proposed design. Fig.~\ref{fig:error_cdf} shows fewer large-error events than RSSI-only \cite{li2026rss}, unweighted TDoA \cite{mandic2024tdoa}, SNR-weighted TDoA \cite{zhu2021nonstationary}, and robust TDoA \cite{abrar2025robust}. Fig.~\ref{fig:rmse_distance} shows that RMSE increases near the anchor-boundary region due to geometric dilution and poorer FIM conditioning. Finally, at $20$ dB, the proposed method achieves about $0.835$ success probability for error below $5$ m, compared with $0.796$, $0.793$, and $0.812$ for unweighted, SNR-weighted, and robust TDoA, respectively. This improvement is useful for threshold-based IoUT localization because it increases the probability of meeting the $5$ m error requirement in multipath environments.

\section{Conclusion}\label{conclusion}
This letter proposed a common-transmitter, multipath-aware TDoA localization framework for acoustic backscatter-enabled IoUT networks. Receiver-side differencing cancels the unknown interrogation time, the forward anchor-to-tag delay, and the tag switching delay, thereby enabling clock-free passive/semi-passive tags. The proposed WNLS estimator combines SNR, timing-noise variance, and RMS delay spread to down-weight unreliable receiver pairs. Results with a noise-only CRLB/FIM reference show gains over RSSI-only, unweighted, SNR-weighted, and robust TDoA. At $20$ dB SNR, RMSE decreases from $4.71$ m to $4.13$ m, while the $5$ m-success probability improves from $0.793$ to $0.835$. Future work includes ray tracing, experimental validation, clock-drift compensation, and moving-node tracking.

\bibliographystyle{ieeetr}
\bibliography{mybibliography}

\end{document}